\documentclass[nohyper,12pt,letterpaper]{JHEP3}
\pdfoutput=1
\usepackage{epsfig}
\usepackage{amsmath}
\usepackage{subfigure}

\newcommand{\be}{\begin{equation}}
\newcommand{\ee}{\end{equation}}
\newcommand{\bea}{\begin{eqnarray}}
\newcommand{\eea}{\end{eqnarray}}
\newcommand{\bwt}{\begin{widetext}}
\newcommand{\ewt}{\end{widetext}}

\def\ie{{\it i.e.~}}

\title{Fast Scramblers and Non-Commutative Gauge Theories}

\author{Mohammad Edalati\\
Department of Physics and Texas Cosmology Center\\ The University of Texas at Austin,
TX 78712.
\\ E-mail: \email{edalati@physics.utexas.edu}}

\author{Willy Fischler\\
Department of Physics and Texas Cosmology Center\\ The University of Texas at Austin,
TX 78712.
\\ E-mail: \email{fischler@physics.utexas.edu}}

\author{Juan F. Pedraza\\
Department of Physics and Texas Cosmology Center\\ The University of Texas at Austin,
TX 78712.
\\ E-mail: \email{jpedraza@physics.utexas.edu}}

\author{Walter Tangarife Garcia\\
Department of Physics and Texas Cosmology Center\\ The University of Texas at Austin,
TX 78712.
\\ E-mail: \email{wtang@physics.utexas.edu}}

\abstract{Fast scramblers are quantum systems which thermalize in a time scale logarithmic in the number of degrees of freedom of the system. Non-locality has been argued to be an essential feature of fast scramblers. We provide evidence in support of the crucial role of non-locality in such systems by considering the approach to thermalization in a (strongly-coupled) high temperature non-commutative gauge theory. We show that non-locality inherent to non-commutative gauge theories does indeed accelerate the rate of dissipation in the heat bath in stark contrast to the slow random walk diffusive behavior prevalent in local field theories.}

\keywords{fast scramblers, horizons, non-commutative}

\preprint{UTTG-06-12\\TCC-008-12}

\begin{document}




\section{\bf Introduction}

It has been known since the days of the ``membrane paradigm" \cite{Thorne:1986iy} that the event horizon of black holes, or more precisely the stretched horizon, is endowed with physical properties from the vantage point of some fiducial observers. This is in contrast with the free infalling observers which do not experience the horizon.  In particular, from the viewpoint of these fiducial observers, a charge falling into the horizon is seen as having its charge smeared over the horizon in a time that grows logarithmically with the Schwarzschild radius of the black hole.

In a remarkable paper, Preskill and Hayden \cite{Hayden:2007cs}  posed a critical challenge to the resolution of the information paradox. In a nutshell, their observation lead to the conclusion that the time scale involved in possibly sharing information between two observes,  one hovering outside the black hole horizon and the other falling through the horizon while carrying information, is dramatically shorter than what was believed possible until then. The fascinating answer to this challenge,  the so-called ``fast scrambling" property of the horizon, was conjectured by Sekino and Susskind in \cite{Sekino:2008he}. This property is tantamount to the extremely rapid spreading of  the charge as we alluded to earlier. The time scale involved in scrambling grows only logarithmically with the entropy. This is a much shorter time scale as a function of the entropy than the time that a particle subject to Brownian motion would take to travel through a distance commensurate with the ``size" of the system \cite{Sekino:2008he}. The time scale in the latter case grows with a positive power of the entropy.

It was argued in \cite{Sekino:2008he} that non-locality has to be one of the ingredients that generate fast scrambling. Local interactions, on the other hand, would lead to slow diffusion as in Brownian motion. Another possibly necessary ingredient to generate the fast scrambling property of the stretched horizon is the presence of strong coupling interactions among the horizon's degrees of freedom \cite{Sekino:2008he}.  

In this paper, we will emulate non-locality and strong coupling by considering a strongly-coupled non-commutative gauge theory at high temperature: four-dimensional non-commutative (supersymmetric) Yang-Mills theory, to be precise.  As the non-commutative system under consideration is strongly coupled, we use the AdS/CFT correspondence as a toolbox for analyzing various thermal aspects of the system. We show that the rate of decay of a disturbance propagating in this thermal medium is parametrically much larger than in the local setting of conventional field theories. This will inevitably lead to rapid thermalization\footnote{It should be noted that the rapid thermalization that we observe here is for the boundary non-commutative thermal bath and does  not seem to be directly connected to the properties of the stretched horizon of the dual black brane background as observed by a fiducial observer hovering outside the horizon.}. It is interesting to note that computations at weak coupling do not exhibit an enhancement of the thermalization rate. 
This could then imply that fast thermalization in non-commutative field theories is a consequence of having both strong coupling and non-commutativity. See
\cite{Susskind:2011ap, Asplund:2011qj, Barbon:2011pn, Lashkari:2011yi} for other works in analyzing the fast scrambling conjecture.

The existence of a connection between the ultraviolet and the infrared in the non-commutative setting has been well studied over the years \cite{Bigatti:1999iz}. One of its manifestation is the large transverse size of any degree of freedom moving with a large momentum $k$, where transversality here is with respect to the direction of $k$. The transverse size grows with $k$ when the momentum in the non-commutative direction is larger than $\theta^{-1/2}$ where $\theta$ is the Moyal area. Indeed, the transverse size scales like $\theta k$. This property is at the heart of the rapid decay of high momenta modes into excitations of the heat bath.

The paper is organized as follows: in the following section we briefly review the gravity dual of a strongly coupled large-$N$ non-commutative gauge theory at finite temperature and then set-up the study of the approach to thermalization in this context.
In section three, we exhibit the computation of the decay rates in the non-commutative heat bath and compare it to the results for the local case of strongly-coupled commutative gauge theories.  More specifically, we compute the decay rates in three cases: once when the non-commutative thermal bath is perturbed by an operator dual to a minimally-coupled massless scalar field in the bulk, then with the operator dual to the dilaton fluctuation of the bulk, and finally with the operator dual to the bulk axion fluctuation.  Furthermore, we compute the decay rates in the regime of parameters where the temperature of the bath is taken to be larger than $\theta^{-1/2}$ along with  the momentum of the modes in the non-commutative plane being greater than the temperature (which necessarily implies that the momentum is also greater than $\theta^{-1/2}$). This is the regime where one expects the deviations from the results of the commutative setting to be more pronounced\footnote{This is due to the fact that at temperatures larger than $\theta^{-1/2}$, the typical momenta in the heat bath is of the order of temperature and hence the transverse size of the modes of the heat bath is large and gets larger as the temperature increases. One therefore expects the non-local effects to be important at those high temperatures.}. We close with some concluding remarks and open questions at the end.

\section{A Gravity Dual For Non-commutative SYM}

In string theory, non-commutative gauge theories can be easily realized on the worldvolume of D-branes with a constant Neveu-Schwarz  $B_{\mu\nu}$ field provided that one takes a special limit to decouple the open and closed string modes \cite{Douglas:1997fm, Ardalan:1998ce,Seiberg:1999vs}. Basically, one scales the string tension to infinity and the closed string metric to zero while keeping the background $B_{\mu\nu}$ field fixed \cite{Seiberg:1999vs}. For much of the discussion in this paper, we focus on the four-dimensional non-commutative Yang-Mills theory at finite temperature. We also take the non-commutativity parameter to be non-zero only in the $(x^2,x^3)$-plane, that is $[x^2,x^3]\sim i \theta$. In the spirit of the AdS/CFT correspondence \cite{Maldacena:1997re}, the dynamics of four-dimensional non-commutative Yang-Mills theory at large $N$ and at strong 't Hooft coupling should be described by a bulk gravitational system. Indeed, a proposal for the dual gravity background was given in \cite{Hashimoto:1999ut,Maldacena:1999mh} which, in the string frame, reads
\begin{align}\label{NCbackground}
ds^2 &=L^2\Big[\hskip-0.03in
-u^2f(u) dt^2+u^2dx_1^2
+u^2h(u)\left(dx_2^2+dx_3^2\right)+ \frac{du^2}{u^2f(u)}+ d\Omega_5^2\Big],\nonumber\\
B_{23} &=L^2a^2 u^4h(u),\qquad\quad
e^{2\Phi} = {\hat g}^2h(u), \\
C_{01} &= L^2 a^2 {\hat g}^{-1}u^4,\qquad
F_{0123u} =4L^4 {\hat g}^{-1}u^3h(u)\nonumber,
\end{align}
where $L^4= 4\pi {\hat g} N {\alpha'}^2$ with $\hat g$ being the value of the string coupling in the IR, and
\begin{align}
h(u)=\frac{1}{1+a^4u^4}, \qquad\qquad f(u)=1-\frac{u_h^4}{u^4},
\end{align}
where $u_h$ is the horizon radius.

The parameter $a$ which appears in the above expressions is related to the non-commutativity parameter $\theta$ of the boundary theory  through $a=\lambda^{1/4}\sqrt{\theta}$. Here, $\lambda=L^4/{\alpha'}^2$ is the 't Hooft coupling of the boundary large-$N$ non-commutative Yang-Mills theory. For $u \ll a^{-1}$,  the background \eqref{NCbackground} goes over to the AdS$_5 \times {\rm S}^5$ solution.  As expected, this observation just reflects the fact that the non-commutative boundary theory goes over to the ordinary commutative Yang-Mills theory at length scales much greater than $\lambda^{1/4}\sqrt{\theta}$. For $u\gg a^{-1}$, on the other hand, the background \eqref{NCbackground} shows significant deviation form the AdS$_5 \times {\rm S}^5$ solution. In particular, the bulk spacetime is no longer asymptotically AdS$_5$. The boundary theory interpretation of this regime just means that the effect of non-commutativity becomes pronounced for length scales which are at the order of, or smaller than, $\lambda^{1/4}\sqrt{\theta}$ \footnote{Since the reliability of the solution \eqref{NCbackground} requires $L^4/{\alpha'}^2=\lambda$ to be large, the effect of non-commutativity in the boundary theory is visible even at large length scales \cite{Maldacena:1999mh}. This should be compared with the weak-coupling regime of the theory where the threshold length scale (beyond which the theory becomes effectively commutative) is roughly at the order of $\sqrt{\theta}$.}. To summarize, the background \eqref{NCbackground} represents a flow in the boundary theory from a UV fixed point which is non-commutative Yang-Mills at large $N$ and large $\lambda$ to an IR fixed point given by the ordinary commutative Yang-Mills (again, at large $N$ and large $\lambda$).

Also, the Hawking temperature of the above solution, which is interpreted as the temperature of the non-commutative boundary theory, is given by $T=u_h/\pi$. Notice that this temperature is the same as the temperature one obtains for the Schwarzschild AdS$_5$ solution, which is dual to a thermal state of four-dimensional commutative Yang-Mills theory.   Indeed, it is easy to show that all the thermodynamic quantities obtained from \eqref{NCbackground} are the same as the ones obtained from the Schwarzschild AdS$_5$ solution \cite{Maldacena:1999mh}.

\section{Approach to Thermalization}

In the context of the AdS/CFT correspondence \cite{Maldacena:1997re, Gubser:1998bc, Witten:1998qj}, turning on the fluctuations of the bulk background translates into deforming the boundary field theory by the operators dual to the bulk fluctuations. In the case where the background is a large black hole, which describes a thermal state in the boundary field theory \cite{Witten:1998zw}, the bulk fluctuations correspond to the perturbations of the thermal bath in the boundary field theory.  Due to the existence of the horizon in the bulk, the fluctuations of the black hole background fall into the horizon. From the perspective of the boundary field theory, the decay of the bulk fluctuations into the horizon translates into the decay of the perturbations of the thermal bath and their eventual thermalization, after some time scale, with the rest of the bath  \cite{Horowitz:1999jd}. In other words, the decay of the bulk fluctuations into the horizon holographically describes the approach to thermal equilibrium in the boundary field theory.

In the linear response theory, the time scale required for a perturbation to thermalize with the rest of the bath can be read off from the imaginary part of the poles in the retarded Green's function of the operator by which the thermal bath has been perturbed. Indeed, the late-time decay is dominated by the imaginary part of the lowest of such poles (in the complex frequency plane). While the computation of  the thermalization time scales is relatively easy in the perturbation theory \cite{Kapusta:2006pm,lebellac}, it is a formidable task when the thermal bath is strongly coupled. Fortunately, this is the regime where the AdS/CFT correspondence is most useful. In fact, the poles in the retarded Green's function of the boundary theory operators are mapped, through the correspondence, to the quasi-normal frequencies of the bulk fluctuations which are dual to those operators, a connection which was first proposed in \cite{Birmingham:2001pj} and later sharpened in \cite{Son:2002sd, Kovtun:2005ev}. See \cite{Berti:2009kk} for a relatively recent review on the quasi-normal modes of various backgrounds (with different asymptotics) from the perspectives of both general relativity and the AdS/CFT correspondence. 

The quasi-normal modes of a perturbation of a gravitational background are defined as the solutions to the equation of motion of that perturbation which are infalling near the horizon and normalizable at the boundary (when the background is asymptotically AdS). The quasi-normal frequencies are then the set of complex frequencies for which such solutions exist. Thus, for strong-coupling thermal baths, the thermalization time scales can simply be obtained by computing the quasi-normal modes of the dual black hole backgrounds. While there are semi-analytic methods for computing special values of the quasi-normal frequencies, their generic values, however, cannot be computed analytically for most backgrounds. In such cases, one can compute them numerically with high precision using a variety of techniques developed over the years in the literature \cite{Berti:2009kk}.

Our goal is to study  the approach to thermalization in a strongly coupled non-commutative thermal bath, which is  dual to the non-extremal background given in \eqref{NCbackground}. In particular, we would like to analyze to what extent the thermalization properties of a non-commutative thermal bath deviate from its commutative counterpart. To do so, we will consider three examples of fluctuations of the dual supergravity background: a minimally-coupled scalar field, fluctuation of the dilaton as well as the fluctuation of the Ramond-Ramond (axion) scalar field. 

\subsection{Minimally-Coupled Scalar}\label{MCS}

We start by considering perhaps the simplest possibility. That is, we perturb the non-extremal background \eqref{NCbackground} by a minimally-coupled massless scalar $\varphi$, which we assume to be dual to some scalar operator $\cal O$ in the non-commutative boundary theory.  Our goal is to calculate the time scale for this perturbation to thermalize with the non-commutative bath.  As we alluded to the above, we need to compute the quasi-normal modes of the minimally-coupled scalar field fluctuations around the background \eqref{NCbackground}. The equation of motion for $\varphi$ (with no dependence on the coordinates of  ${\rm S}^5$) reads
\begin{align}\label{SEoM}
\frac{e^{2\Phi}}{\sqrt{-g}}\,\partial_\mu\left(\sqrt{-g}\,e^{-2\Phi}g^{\mu\nu}\partial_\nu\varphi\right)=0,
\end{align}
which, in the non-extremal background \eqref{NCbackground}, takes the form
\begin{align}\label{SEoMPositionSpace}
0=&\,\partial_u\left[u^5f(u)\partial_u\varphi\right]+u\Big[-\frac{1}{f(u)}\partial_t^2+\partial_{x^1}^2+\frac{1}{h(u)}\left(\partial_{x^2}^2+\partial_{x^3}^2\right)\Big]\varphi.
\end{align}

We emphasize that the minimally-coupled massless scalar that we are studying here is not related to the metric fluctuations of the background \eqref{NCbackground} such as $\delta g^{t}_{\,x^1}$. As shown in \cite{Maldacena:1999mh}, the equation of motion for $\delta g^{t}_{\,x^1}$ (with no dependence on ${\rm S}^5$) satifies the same equation as in \eqref{SEoM} provided that $\delta g^{t}_{\,x^1}$ depends neither on $t$ nor on $x^1$ directions. It is not clear to us whether the equation of motion for any generic fluctuation of the background \eqref{NCbackground}, or their  gauge invariant combinations,  could be given by the equation \eqref{SEoMPositionSpace}. As such, the analysis in this section should be regarded as a warm-up exercise.  Nevertheless, we believe that the lesson we learn by analyzing the case of a minimally-coupled massless scalar will prove useful when we consider, in the subsequent section,  some genuine fluctuations of the background such as the dilaton and axion fluctuations. 

To solve the equation \eqref{SEoMPositionSpace}, we go to momentum space by Fourier transforming $\varphi(u,t,\vec x)$:
\begin{align}
\varphi(u,t,\vec x)\sim e^{-i\omega t +i{\vec k}.{\vec x}}\varphi(u; \omega, \vec k).
\end{align}
To avoid clutter in the notation, we have denoted the Fourier modes of the scalar field also by $\varphi$. The distinction should be clear from the arguments of $\varphi$, or the context in which it is being used. In order to see any non-trivial effect on the behavior of the modes  $\varphi(u; \omega, \vec k)$ due to the existence of a non-zero Neveu-Schwarz $B_{\mu\nu}$ field in the bulk, especially in the asymptotic $u\to\infty$ region where the behavior of $\varphi(u\to\infty; \omega, \vec k)$ holographically encodes information about the source and the condensate of the dual operator ${\cal O}(\omega,\vec k)$, one should consider the modes with a non-vanishing momentum in the $(x^2, x^3)$ plane. Without loss of generality, we take $k_2 = 0$ and $k_3\neq 0$ and, for simplicity, we also consider the modes with $k_1=0$. The equation of motion \eqref{SEoMPositionSpace} then becomes
\begin{align}\label{SEoMMomentumSpace}
\partial_u\left[u^5f(u)\partial_u\varphi\right]+u\left[\frac{\omega^2}{f(u)}- \frac{k^2}{h(u)}\right]\varphi=0,
\end{align}
where $k\equiv k_3$. 

Some comments on the boundary theory operator dual to $\varphi(u,t,\vec x)$ are in order here. Normally, in the AdS/CFT correspondence, one associates a gauge invariant operator ${\cal O}(t,\vec x)$ in the boundary theory to the bulk fluctuation $\varphi(u,t,\vec x)$ in the sense that near the boundary of the spacetime the non-normalizable mode of $\varphi(u\to\infty,t,\vec x)$ sources the dual operator ${\cal O}(t,\vec x)$ while its normalizable mode gives the vacuum expectation value of    ${\cal O}(t,\vec x)$. The situation is less clear when the boundary gauge theory is non-commutative. This is  partly due to the fact that in non-commutative gauge theories, non-commutativity of the space mixes with the gauge transformations and, as such, there are no gauge invariant operators in position space. In momentum space, however, one can construct gauge invariant operators ${\cal O} (k^\mu)$ by smearing the gauge covariant operators ${\cal O}(x^\mu)$ transforming in the adjoint over an open Wilson line $W(x,C)$ of definite size according to \cite{Ishibashi:1999hs, Gross:2000ba,Das:2000b} 
\begin{align}\label{wline}
{\cal O}(k) = \int d^4x\,{\cal O} (x)\star W(x,C) \star e^{i k \cdot x}, 
\end{align}
where $\star$ denotes the Moyal product. The expression in \eqref{wline} reduces to the corresponding ordinary gauge invariant local operators in the IR. In the UV, on the other hand, this very simple modification, has a drastic consequence for the behavior of the correlation functions of the gauge invariant operators ${\cal O}(k)$. 

In the following we assume that the bulk fluctuation $\varphi(u; \omega, \vec k)$ is dual to a gauge invariant operator ${\cal O}(\omega,\vec k)$ of the form \eqref{wline}  in the sense that in the boundary non-commutative gauge theory there exists a coupling of the following form in the momentum space, 
\begin{align}\label{BoundaryCoupling}
\int d\omega d^3k\,\varphi_0(-\omega,-\vec k){\cal O}(\omega, \vec k),
\end{align}
where $\varphi_0(\omega,\vec k)$ is the source term which is determined from the asymptotic non-normalizable behavior of $\varphi(u,\omega,\vec k)$ given some suitable boundary condition near the horizon. 

To obtain the retarded Green's function of the dual operator ${\cal O}(\omega,k)$, equation \eqref{SEoMMomentumSpace} should then be solved with infalling boundary condition at the horizon \cite{Son:2002sd}. Close to the horizon, the solution behaves as
\begin{align}
\varphi(u)\sim f(u)^{\pm i\omega/(4\pi T)},\qquad u\to u_h,
\end{align}
where the + and $-$ signs represent outgoing and infalling waves,  respectively.  So, we discard the solution with the + sign. Near the boundary $u=\Lambda \to \infty$, the equation \eqref{SEoMMomentumSpace} can be conveniently cast into a Mathieu differential equation. The asymptotic behavior of the solution then takes the form
\begin{align}\label{AsymptoticOfScalar}
\varphi(u)\sim A(\omega, k)\, u^{-5/2} e^{k a^2 u} + B(\omega, k)\,u^{-5/2} e^{-k a^2 u}.
\end{align}
The first term in the above equation blows up as the cutoff $\Lambda\to \infty$. So, in the spirit of the AdS/CFT correspondence, the source for  the operator dual to $\varphi$ in the boundary  theory should be read off from that term. However, extracting the source of the dual operator from \eqref{AsymptoticOfScalar} is ambiguous. The ambiguity
arises because, as seen from \eqref{AsymptoticOfScalar}, for any cutoff $\Lambda$ the relation between the would-be source $A(\omega, k)$ and the asymptotic behavior of $\varphi$ involves a function of $k$, namely $e^{ka^2\Lambda}$. As a result, the normalization of the dual operator, and hence the normalization of the correlation functions, in the boundary theory is not fixed\footnote{This is in contrast to the usual examples in the AdS/CFT correspondence where the background is asymptotically AdS and the relation between the asymptotic behavior of a bulk fluctuation and the source of the corresponding dual operator does not depend on $k$. Indeed, such a relation is fixed by the conformal symmetry of the boundary theory.}. This feature, which also arises in \cite{Aharony:1998ub, Minwalla:1999xi}, can ultimately be attributed to the non-local nature of the boundary theory in the UV.  As advocated in \cite{Maldacena:1999mh} (see also \cite{Das:2000fua}), a reasonable way forward is to define the source for the dual operator from the asymptotic behavior of the bulk fluctuation in such a way that the correlation functions of the operator reproduce the ordinary commutative results in the IR. Nevertheless, regardless of how the normalization of the boundary theory operator is fixed, the ratios of the correlation functions are unambiguous \cite{Maldacena:1999mh, Gross:2000ba}. Also, note that the overall $k$-dependent normalization should not affect the poles (in the complex frequency plane) of the retarded Green's function of the dual operator. 

In order to find the poles in the retarded Green's function of the dual operator ${\cal O}(\omega,k)$, we numerically solve \eqref{SEoMMomentumSpace} with infalling boundary condition near the horizon and normalizable boundary condition near  the boundary\footnote{The normalizable boundary condition near the boundary seems to be a natural choice given that the potential term in equation \eqref{SEoMMomentumSpace}, once written in the form of a Schr\"{o}dinger equation, blows up near the boundary, see the equation \eqref{SchrodingerEq} in the appendix. This is similar to the case where the background is asymptotically AdS where the potential term in the equation of motion for a minimally-coupled scalar also blows up at the boundary of the spacetime.}. In other words, we compute the quasi-normal frequencies of $\varphi$ in the non-extremal background \eqref{NCbackground}. To proceed, we find it convenient to define a new radial coordinate $z=u_h/u$. In terms of this new radial coordinate, the horizon is now at $z=1$ while $z\to 0$ is the asymptotic boundary of the spacetime. Using the relation $u_h=\pi T$ with $T$ being the Hawking temperature of the background, the equation of motion \eqref{SEoMMomentumSpace}, expressed in terms of the new radial coordinate, takes the form
\begin{align}\label{SEoMzCoordinate}
0&= f(z)z^2\varphi''(z)+\left[zf'(z)-3f(z)\right]z\varphi'(z)+\frac{z^2}{\pi^2T^2}\left[\frac{\omega^2}{f(z)}- \frac{k^2}{h(z)}\right]\varphi(z),
\end{align}
where the primes denote derivatives with respect to $z$.  
Equation (\ref{SEoMzCoordinate}) has three (real-valued) regular singular points at $z=-1,1,\infty$ and one irregular singularity at $z=0$, thus we must proceed with some care. To numerically compute the quasi-normal frequencies, we employ a method first developed in \cite{Leaver:1990zz}.  That is, we pull out from $\varphi$ its infalling behavior near the horizon as well as its leading normalizable behavior near the boundary and define a new function $\tilde \varphi$ according to
\begin{align}\label{sclr-sltn}
\varphi(z;\omega,k)= f(z)^{-i\omega/(4\pi T)} z^{5/2}e^{-\pi k a^2 T/z}\tilde\varphi(z;\omega,k).
\end{align}
The equation for $\tilde\varphi$, obtained from \eqref{SEoMzCoordinate}, does not have any singular points in the interval $z\in[1,0]$. Hence, we can approximate $\tilde\varphi$ in that  interval by a series expansion, with $M$ terms in it,  around a point inside the interval, say $z_0=1/2$ (so that the radius of convergence of the series covers the two endpoints of the interval):
\begin{align}\label{SeriesTildeVarphi}
\tilde\varphi(z;\omega,k)=\sum_{m=0}^M a_m(\omega,k) \left(z-\frac{1}{2}\right)^m.
\end{align}
The equation for $\tilde\varphi$ can then be cast in the following form
\begin{align}
\sum_{m=0}^M A_{mn}(\omega,k) a_n(\omega,k)=0,
\end{align}
which is a matrix equation describing $M+1$ linear equations for $M+1$ unknowns $a_n(\omega,k)$. For any fixed $k$, the quasi-normal frequencies, denoted by $\omega_n$, are then obtained by solving the following polynomial equation in $\omega$:
\begin{align}\label{det}
\det A_{mn}(\omega,k)=0.
\end{align}
As we increase the size of the matrix $A$ defined above, the poles are found to move around in the complex frequency plane slightly. However, depending on the desired precision, they converge for a large but finite value of $M$. For concreteness, we ran all our computations with $M=300$, for which we found a confidence level up to fourteen significant digits for the imaginary part of the first mode.

Now, since the boundary field theory under consideration is non-commutative, there is another scale in the problem besides the temperature, namely, the length scale $\sqrt{\theta}$ (or rather $\lambda^{1/4}\sqrt{\theta}$ for our strong coupling setting).  
As such, the poles in the retarded Green's function of the operator ${\cal O}(\omega, k)$, or equivalently, the quasi-normal frequencies of the dual bulk fluctuation, depend on $k$, $T$ and $1/\sqrt{\theta}$ and can be conveniently parametrized as
\begin{align}
\omega_n(k,T, \theta) = T f_n\Big(\frac{k}{T}, T\sqrt{\theta}\Big).
\end{align}
The index $n\geq 0$ which is sometime referred to as overtone number denotes the number of poles. For instance, $n=0$ is assigned to the pole with the smallest magnitude. Aside from special regions in the parameter space of $k/T$ and $T\sqrt{\theta}$,  where one can determine the functions $f_n$ semi-analytically, $f_n$s should, in general, be determined numerically. As we alluded to earlier, one expects the deviations from the  commutative results to be more pronounced in the region of  parameter space with $T\gg \theta^{-1/2}$ and $k \gg T^{-1}$ which also implies $k\gg \theta^{-1/2}$. As we demonstrate below, this expectation is confirmed in our numerical computations.

\begin{figure*}[h]
\centering

\subfigure[ ]{\hskip -0.2in \includegraphics[width=57mm]{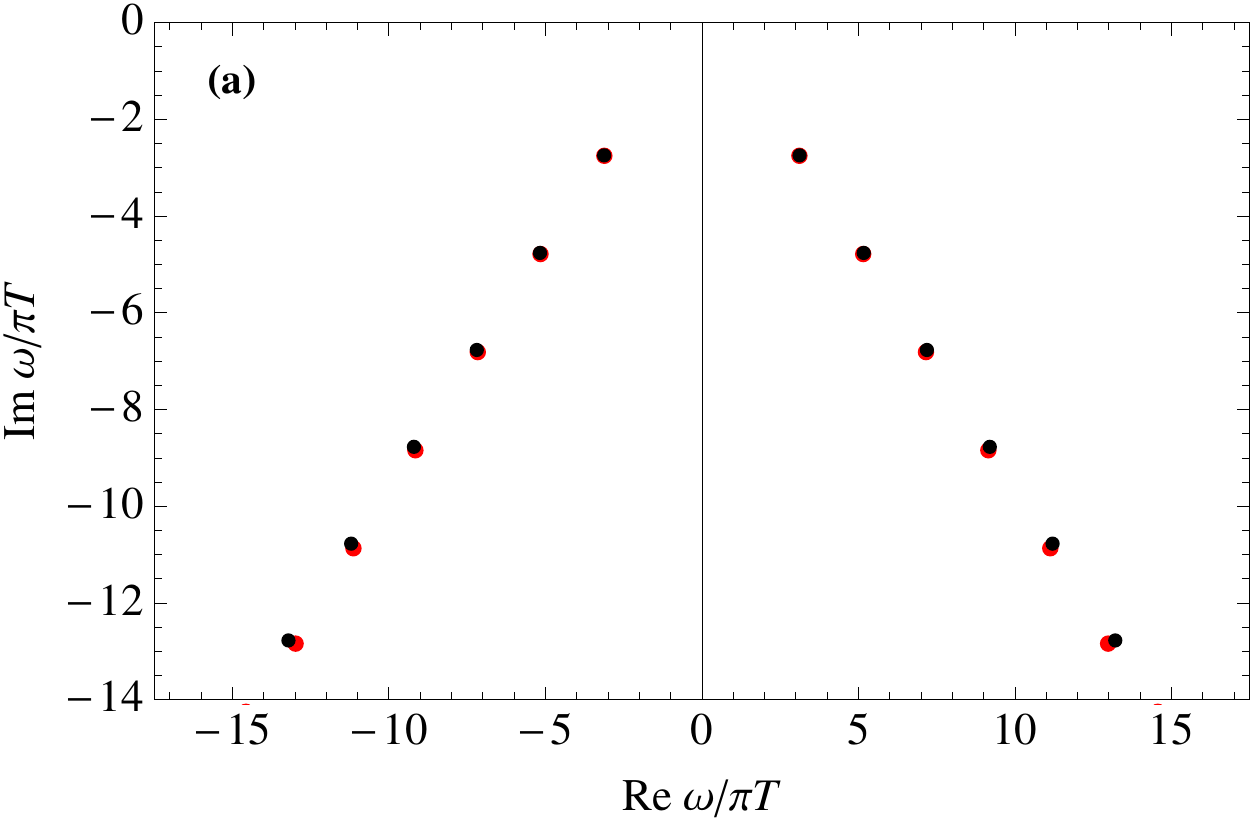}\hskip 0.21in}
\hskip 0.1in
\subfigure[ ]{\hskip -0.26in\includegraphics[width=57mm]{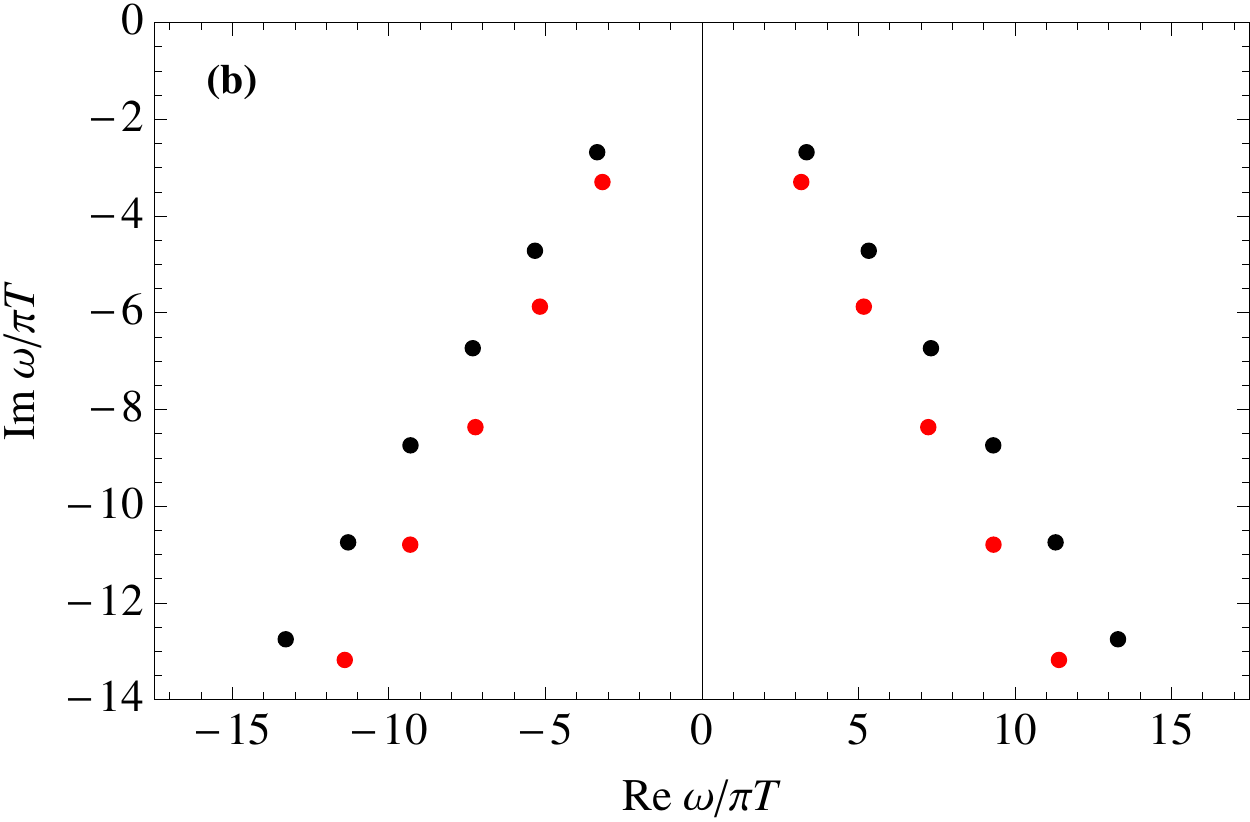}}
\hskip 0.12in
\subfigure[ ]{\hskip -0.26in\includegraphics[width=57mm]{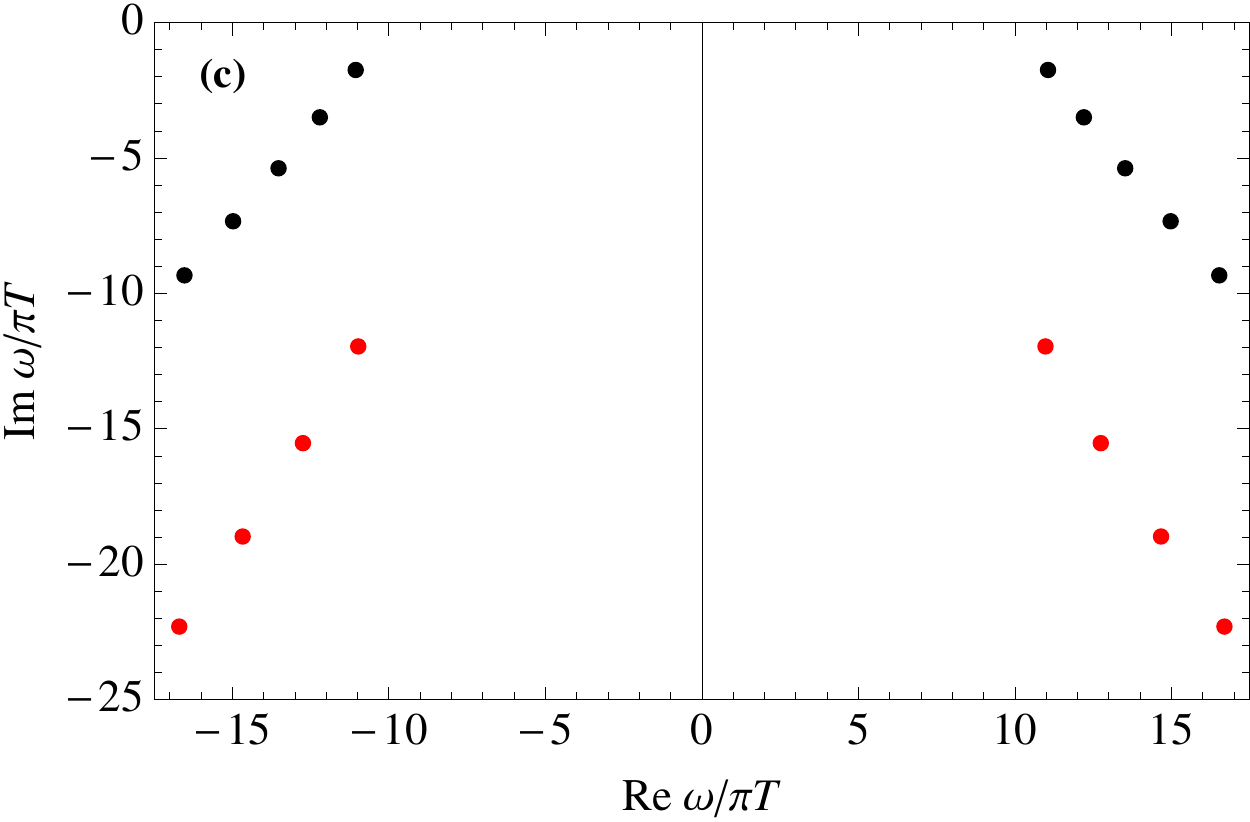}}
\hskip 0.1in
\caption{\label{QNMs}\footnotesize{The red and black dots show, respectively, the location in the complex frequency plane of the first few quasi-normal frequencies of a minimally-coupled massless scalar on the non-extremal background \eqref{NCbackground} and the AdS$_5$ Schwarzschild background  for (a) $k=\pi T/10$, (b) $k=\pi T$ and (c) $k=10\pi T$, where we have kept fixed $\pi a T=1$. The plots have been generated with $M = 300$.} }
\end{figure*}

The plots (a), (b) and (c) in Figure \ref{QNMs} show, respectively, the location of the first few quasi-normal frequencies  for $k=\pi T/10$,  $k=\pi T$ and $k=10\pi T$ for a fixed value of $\pi aT=1$. The red dots show the location of the quasi-normal frequencies of $\varphi$ on the non-extremal background \eqref{NCbackground} while the black dots represent the quasi-normal frequencies of $\varphi$ on the AdS$_5$ Schwarzschild background. Note that even in the presence of a non-zero non-commutativity parameter, the poles are all located in the lower half of the complex frequency plane. 
Indeed, one can easily show that regardless of the range of parameters no poles will move to the upper half plane, indicating the stability of the system. See the appendix for a proof.  

\begin{figure}
\centering
\hskip -0.12in \includegraphics[width=65mm]{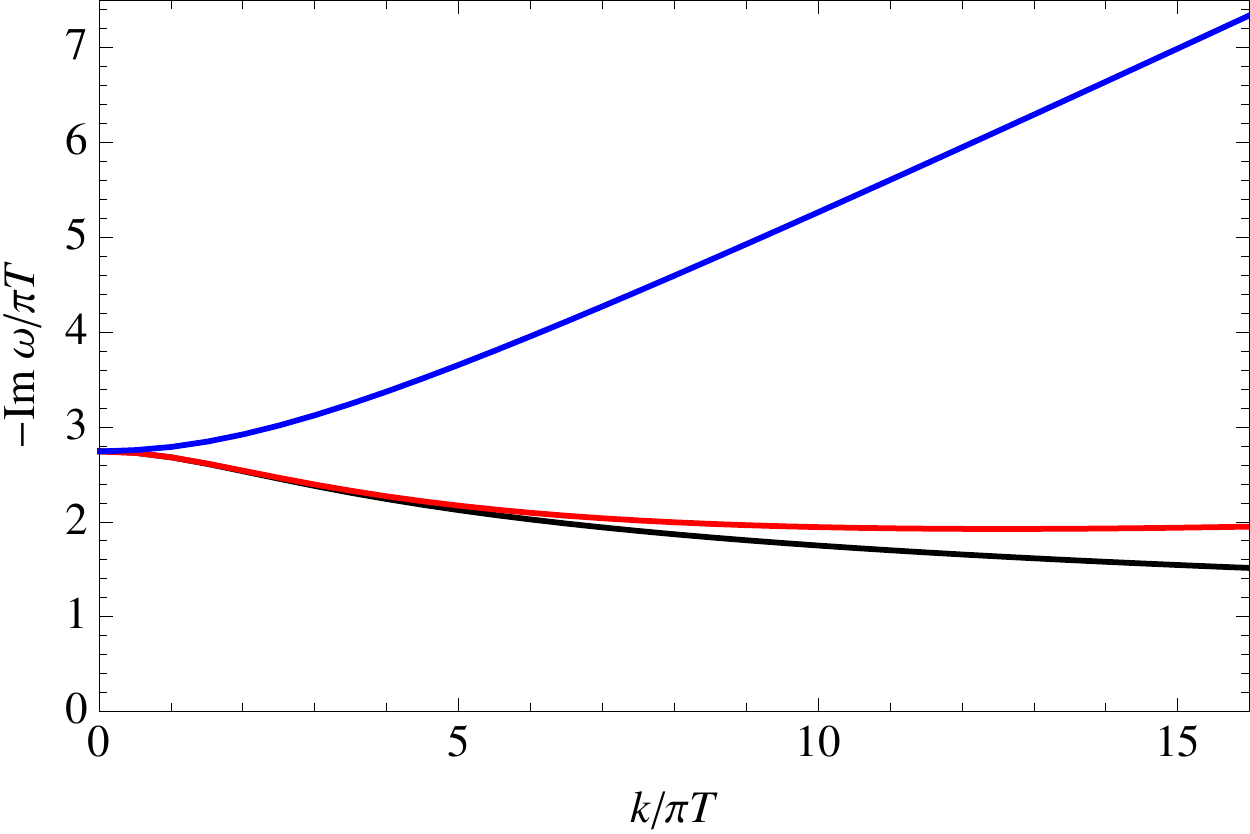}
\hskip 0.1in
\caption{\label{Dispersion}\footnotesize{The behavior of the (negative of the) imaginary part of $\omega/T$ as a function of 
$k/T$ for the dominant quasi-normal frequency in the commutative case ($\theta=0$) shown by the back curve, and in the non-commutative cases with $\pi aT=1/5$ and $\pi aT=3/5$ which are shown by the red and blue curves, respectively. } }
\end{figure}
\begin{figure*}
\centering
\subfigure[ ]{\hskip -0.2in  \includegraphics[width=57mm]{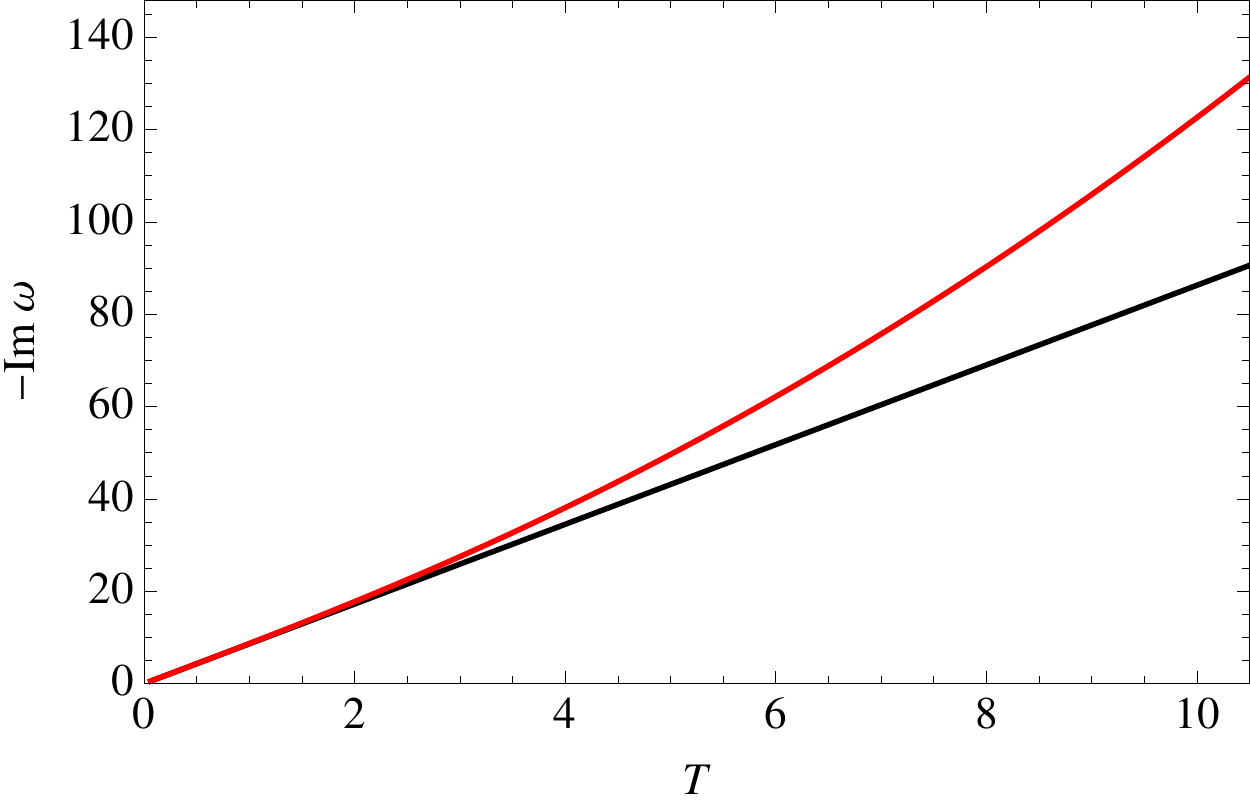}\hskip 0.2in}
\hskip 0.1in
\subfigure[ ]{\hskip -0.3in \includegraphics[width=57mm]{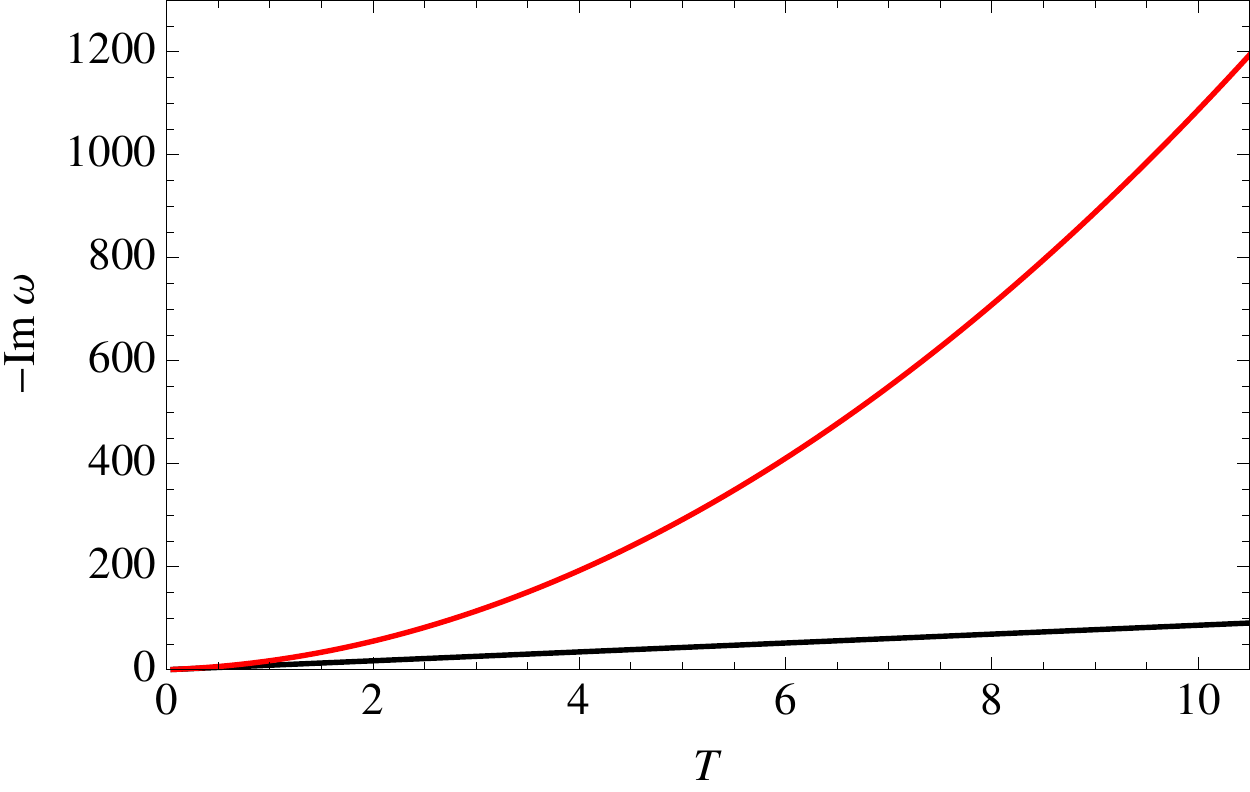}}
\hskip 0.1in
\subfigure[ ]{\hskip -0.3in\includegraphics[width=58mm]{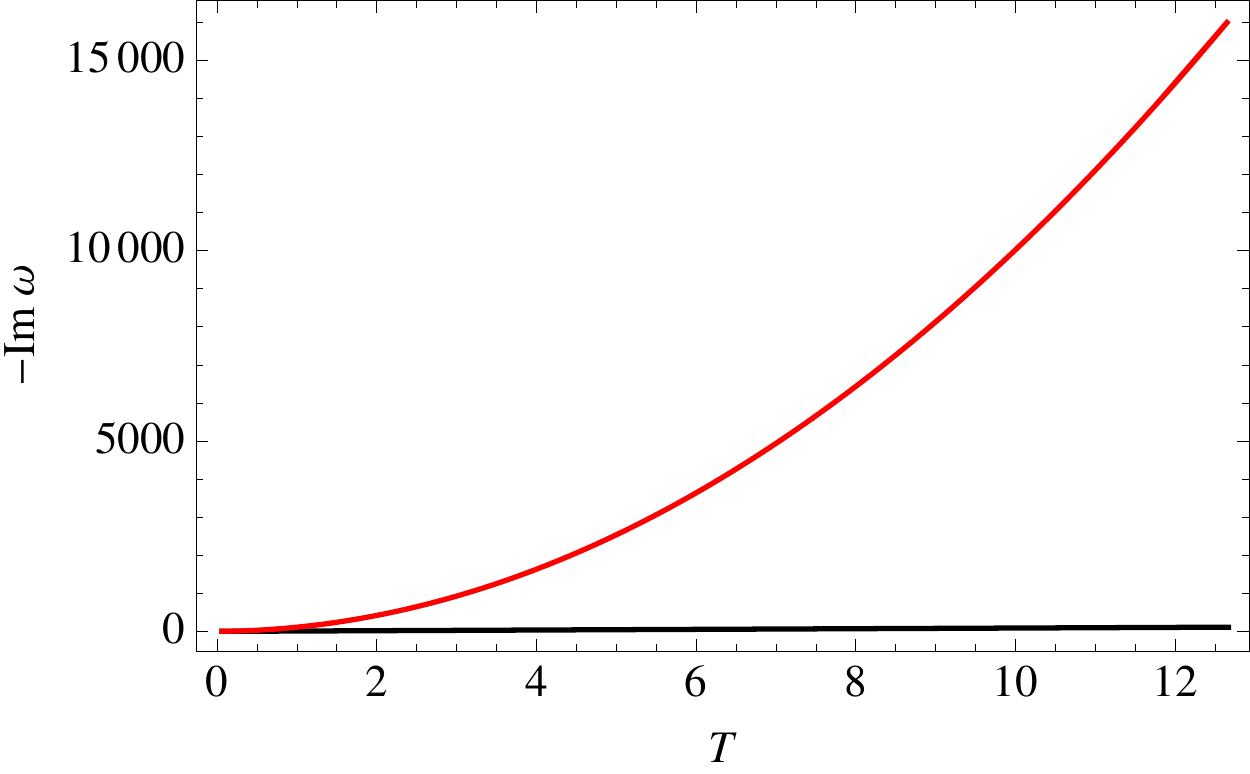}}
\vskip -0.1in
\caption{\label{ImOmegavsT}\footnotesize{The behavior of the (negative of the) imaginary part of the dominant quasi-normal frequency as a function of temperature for $k=5/100$ (a), $k=1$ (b) and $k=10$ (c) in units of  $1/\sqrt{\theta}$. The red curves show the behavior for the non-commutative boundary theory while the black curves show the behavior of the same mode in the commutative theory ($\theta = 0$).} }
\end{figure*}

Focusing on the dominant quasi-normal frequency  (the one with the largest imaginary part), Figure \ref{Dispersion} shows the imaginary part of $\omega/T$ as a function $k/T$, with $T\sqrt{\theta}$ held fixed, for that  mode for both non-commutative and commutative cases. We have also plotted the behavior of its imaginary part versus temperature for $k=5/100$,  $k=1$ and $k=10$ (in units of $a^{-1}$) where we have fixed $a=\lambda^{1/4}\sqrt{\theta} = 1$. Those results are shown in Figure \ref{ImOmegavsT}. In each plot, the red curve shows the non-commutative result while the black curve shows the commutative result ($\theta = 0$) for the same value of momentum $k$. Fitting those curves with a simple power law ${\rm Im}\,\omega_0 = -\alpha\,T^\gamma$ in the regime of $T\sqrt{\theta}\gg1$, we obtain the results reported in the Table I.
\begin{table}[h]
\label{table-mcs}
\centering
\begin{tabular}{|c|c|c|c|c|}
\hline
 \multicolumn{1}{|c}{} & \multicolumn{2}{c}{Commutative}&
\multicolumn{2}{c|}{Non-commutative} \\
 \hline
$k$ & $\alpha$ & $\gamma$ & $\alpha$ & $\gamma$ \\
\hline
$0.05$ & $5.941$ & $\,1.318\,$ & $\,\, 8.629\,\,$ & $1.000$ \\
\hline
$1.00$ & $13.398$ & $\,1.910\,$ & $\,\, 8.623\,\,$ & $1.000$ \\
\hline
$10.00$ & $\,103.64\,$ & $\,1.986\,$ & $\,\, 8.623 \,\,$ & $1.000$ \\
\hline
\end{tabular}
\caption{\footnotesize{Fits of the form ${\rm Im}\,\omega=-\alpha T^\gamma$ for the curves shown in Figure \ref{ImOmegavsT} in the regime of $T\sqrt{\theta}\gg1$.}}
\end{table}

From our numerical results above one can then conclude that in the regime of parameters with both   $T\sqrt{\theta}\gg 1$ and $k\sqrt{\theta}\gg 1$, the decay rate of a mode with momentum $k$ into the non-commutative bath is parametrically faster compared to the decay of the same mode into an ordinary commutative bath. In the regime of high temperatures (as compared to $\theta^{-1/2}$), the non-local effects in non-commutative thermal baths are due to the large transverse size of the typical modes in the bath. These modes provide for a rapid dissipation of the high energy modes injected into the non-commutative bath.  In the case of a black hole or deSitter space,  the event horizon appears to an observer who is hovering outside the horizon increasingly hotter as the observer gets closer to the horizon. Our analysis does show that at increasingly higher temperatures the rate of dissipation increases parametrically much faster in the non-commutative case. This appears consistent with the fast scrambling property of horizons.

\subsection{Dilaton and Axion Fluctuations}

Similar to our analysis in the previous section, we consider in this section the fluctuations of the dilaton and the axion in the background \eqref{NCbackground} and analyze the approach to thermalization of the boundary non-commutaive bath when it gets perturbed by their dual operators.

To start, consider the bosonic part of the action for type IIB supergravity. In the Einstein frame, $ds^2_E=\sqrt{\hat g}\,e^{-\Phi/2}ds_{\rm str}^2$, we can write it as
\begin{align}\label{actIIB}
S&=\frac{1}{2\kappa^2} \int d^{10}x \sqrt{-g}\,{\cal R} - \frac{1}{4\kappa^2} \int \Big( d\Phi \wedge *d\Phi + e^{2\Phi} dC \wedge *dC \\
&+{\hat g} e^{-\Phi} H_3 \wedge * H_3+{\hat g} e^{\Phi} \tilde F_3 \wedge * \tilde F_{3}
+ \frac{1}{2} {\hat g}^2\tilde F_{5} \wedge * \tilde F_{5}
+ {\hat g}^2 C_{4} \wedge H_{3} \wedge F_{3}\Big) \nonumber,
\end{align}
where
\begin{align}\label{fstrens}
\tilde F_{3}&=F_{3} - C H_{3},\qquad\qquad  F_{3}= d C_{2},\nonumber\\
\tilde F_{5}&= F_{5} - C_{2} \wedge H_{3},\qquad\,\, F_{5}= d C_{4}\ .
\end{align}
The type IIB equations of motion can be deduced from (\ref{actIIB}), supplemented by the self-duality condition
\begin{align}
* \tilde F_5 = \tilde F_5\ .
\end{align}
In particular, the background (\ref{NCbackground}) is a solution to the equations of motion obtained from the above action. 

We now proceed by considering the fluctuations of the dilaton and axion around the background \eqref{NCbackground}. As we alluded to earlier, whereas there are no dual gauge invariant local operators in position space, it is possible to write down the linearized coupling of a supergravity field to the noncommutative Yang-Mills modes in momentum space as shown in (\ref{BoundaryCoupling}), with ${\cal O}(k)$ being a gauge invariant operator of the type given in (\ref{wline}) (see, for example, \cite{liu1,liu2,das3}). For the dilaton and the axion, we can think of these operators as the non-commutative gauge invariant generalizations of ${\rm Tr}\, F^2$ and ${\rm Tr}\, F\tilde{F}$, respectively.

The equations of motion that result from the above action for the dilaton $\Phi$ and the axion $C$  are \cite{schwarz}
\begin{align}
&\nabla^2\Phi= e^{2\Phi} \partial_\mu C \partial^\mu C-\frac{ {\hat g}e^{-\Phi}}{12}H_{\mu\nu\rho} H^{\mu\nu\rho}+\frac{{\hat g}e^{\Phi}}{12} \tilde F_{\mu\nu\rho} \tilde F^{\mu\nu\rho},\nonumber\\
&\nabla^\mu( e^{2\Phi} \partial_\mu C) = \frac{{\hat g}e^{\Phi} }{6}
H_{\mu\nu\rho} \tilde F^{\mu\nu\rho},
\end{align}
where it is understood that the indices $\mu,\nu,\ldots$ run in ten dimensions. 

Now, considering small perturbations as defined by $\Phi=\Phi_0+\phi$ and $C=C_0+\chi$, where $\Phi_0$ and $C_0=0$ are the background values given in \eqref{NCbackground}, we get the following differential equations in momentum space (after linearizing):
\begin{align}\label{DE1}
0&= f(z)z^2\phi''(z)+\left[zf'(z)-3f(z)\right]z\,\phi'(z)+\frac{f(z)z^2}{\pi^2T^2}\left[\frac{\omega^2}{f(z)^2}- \frac{k^2}{f(z)\,h(z)}\right.\nonumber\\
&\left. +\frac{16\pi^6T^6\theta^2}{z^6\,f(z)}\left(1+2h(z)\right)+\frac{h'(z)^2}{h(z)^2}\left(\frac{\pi^6T^6\, \theta^2}{z^4}-\frac{\pi^2T^2}{2} \right)+\frac{h''(z)}{2h(z)} \right. \nonumber \\
& \left. +\frac{f'(z)\,h'(z)}{2\,f(z)\,h(z)} -\frac{4\,h'(z)}{h(z)}\left(\frac{2\,\pi^6T^6 \theta^2}{z^5}+\frac{\pi^2T^2}{z} \right)\right]\phi(z),
\end{align}
and
\begin{align}\label{AxionEq}
0&= f(z)z^2\chi''(z)+\left[zf'(z)+\frac{zh'(z)\,f(z)}{h(z)} -3f(z)\right]z\chi'(z)+\frac{z^2}{\pi^2T^2}\left[\frac{\omega^2}{f(z)}- \frac{k^2}{h(z)}\right]\chi(z),
\end{align}
where, as before, we defined $z\equiv u_h/u$ and $k\equiv k_3$.

Close to the horizon, the solutions to both equations behave as
\begin{align}
\phi,\chi\sim f(z)^{\pm i\omega/(4\pi T)},
\end{align}
where the + and $-$ signs represent outgoing and ingoing waves, respectively.  Thus, we discard the solutions with the + sign. Near the boundary, and for non-vanishing momentum, the solutions take the form
\begin{align}\label{AsymptoticOfdilaton}
\phi,\chi\sim A(\omega, k)\, z^{\lambda} e^{k \theta\,\pi\, T/z} + B(\omega, k)\,z^{\lambda} e^{-k \theta\,\pi\, T/z},
\end{align}
where $\lambda=13/2$ for the dilaton and $\lambda=1/2$ for the axion. Notice that their asymptotic behavior is similar to that of a minimally-coupled scalar field as given in \eqref{AsymptoticOfScalar}, although with a different characteristic exponent for $z$. The first term above blows up near the boundary so we take it to be proportional to the source for the boundary theory dual operator. Equations (\ref{DE1}) and (\ref{AxionEq}) have also three real-valued regular singular points at $z=-1,1,\infty$ and one irregular singularity at $z=0$\footnote{For $k=0$ the point $z=0$ is regular in both cases. We do not consider this case here.}. The quasi-normal frequencies of the dilaton and axion fluctuations in the background \eqref{NCbackground} can be computed using the same method we employed in the previous section for the case of a minimally-coupled massless scalar field. 

\begin{figure*}
\centering
\hskip -0.12in 
\subfigure[ ]{\hskip -0.2in \includegraphics[width=57mm]{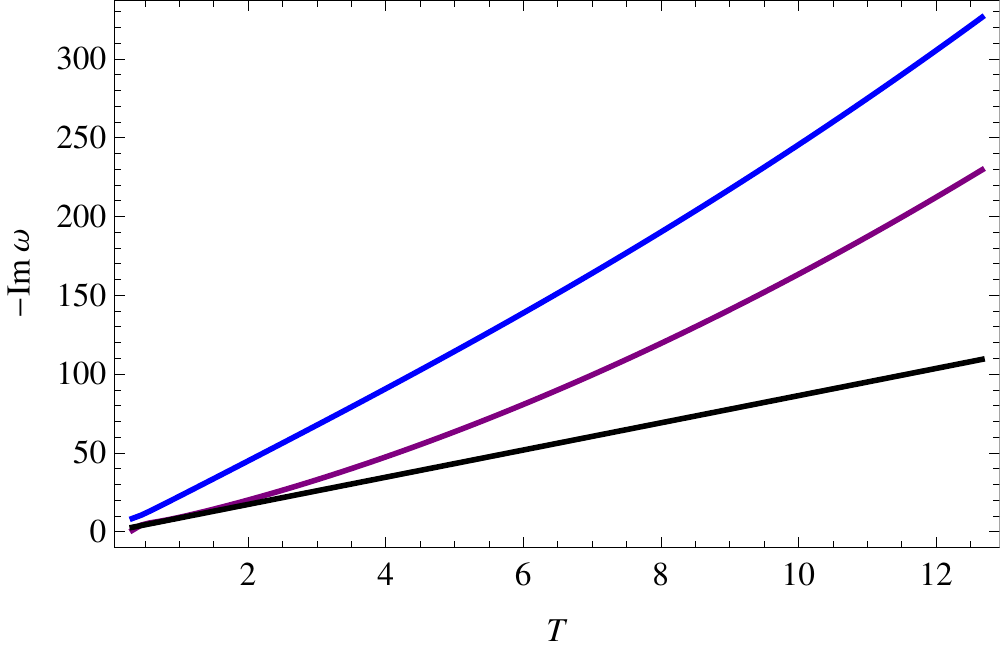} \hskip 0.2in}
\hskip 0.2in
\subfigure[ ]{\hskip -0.3in \includegraphics[width=57mm]{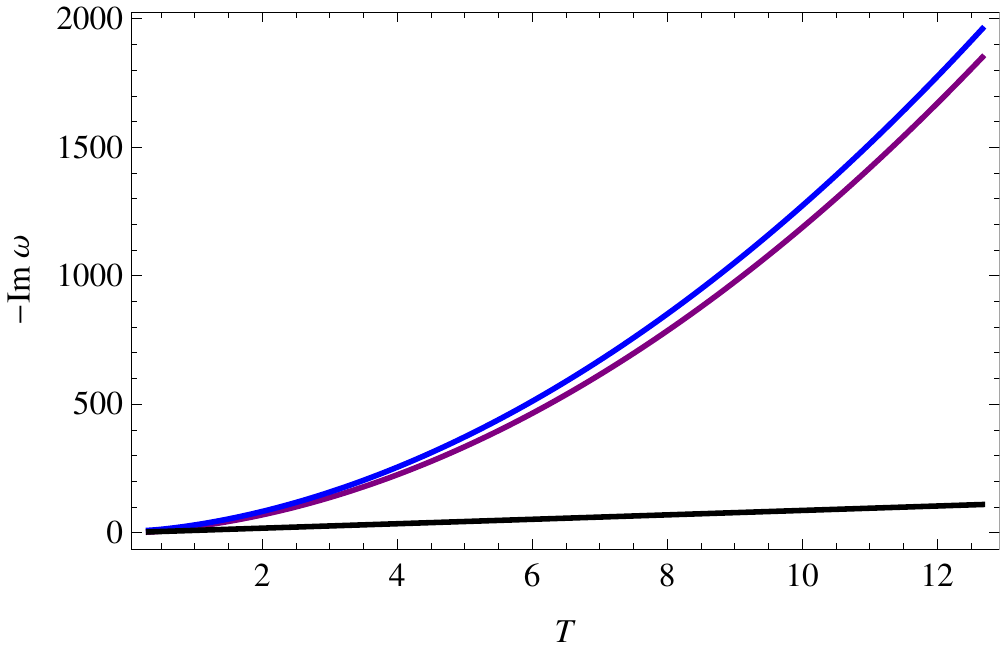}}
\hskip 0.2in
\subfigure[ ]{\hskip -0.3in \includegraphics[width=57mm]{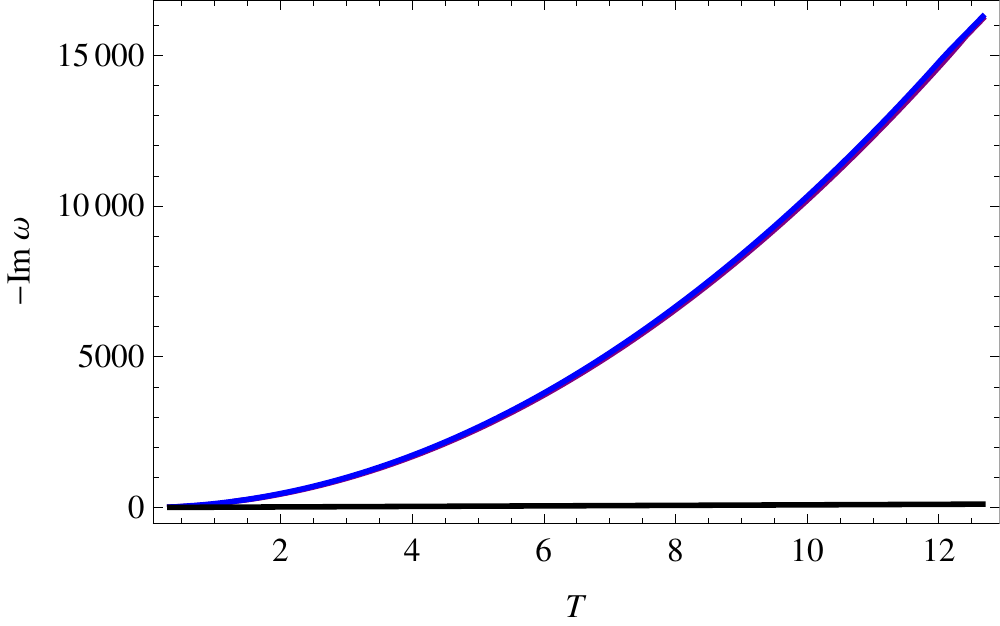}\hskip 0.17in}
\vskip 0.1in
\caption{\label{ImOmegavsT2}\footnotesize{The behavior of the (negative of the) imaginary part of the dominant quasi-normal frequency as a function of temperature for $k=5/100$ (a), $k=1$ (b) and $k=10$ (c) in units of  $1/\sqrt{\theta}$. The blue curves show the behavior for the dilaton fluctuation whereas the purple ones represent the axion fluctuation. For completeness we are also plotting in black the behavior of the same mode for the commutative case. Each point on the plot has been generated with $M = 300$.} }
\end{figure*}

In Figure \ref{ImOmegavsT2} we plotted the behavior of the imaginary part of the dominant quasi-normal frequency as a function of the temperature for $k=5/100$,  $k=1$ and $k=10$ (in units of $a^{-1}$) where we fixed $a=\lambda^{1/4}\sqrt{\theta}=1$. In each plot, the blue curve shows the result for the dilaton fluctuation while the purple one corresponds to the axion fluctuation. For comparison purposes, we also plotted the commutative result  ($\theta = 0$) in black which is same result as for the minimally-coupled massless scalar field in the commutative case\footnote{This is because when the background is AdS (Schwarzschid black hole), the equations of motion for the dilaton, the axion and a minimally-coupled massless scalar are all the same.}. In this case, the results for the best fits are given in the Table II.

\begin{table}
\label{table-dilaton}
\centering
\begin{tabular}{|c|c|c|c|c|}
\hline
 \multicolumn{1}{|c}{} & \multicolumn{2}{c}{Dilaton}&
\multicolumn{2}{c|}{Axion} \\
 \hline
$k$ & $\alpha$ & $\gamma$ & $\alpha$ & $\gamma$ \\
\hline
$0.05$ & $19.188$ & $\,1.1094\,$ & $\,\, 6.9056\,\,$ & $1.3758$ \\
\hline
$1.00$ & $20.691$ & $\,1.7903\,$ & $\,\, 17.047\,\,$ & $1.8442$ \\
\hline
$10.00$ & $\,114.31\,$ & $\,1.9552\,$ & $\,\, 109.14 \,\,$ & $1.9703$ \\
\hline
\end{tabular}
\caption{\footnotesize{Fits of the form ${\rm Im}\,\omega=-\alpha T^\gamma$  for the blue and purple curves shown in Figure \ref{ImOmegavsT2} in the regime of $T\gg\theta^{-1/2}$.}}
\end{table}

These results support the main conclusion found in the previous section: in the regimes at which non-local effects become important, \ie $T\gg \theta^{-1/2}$ and $k\gg \theta^{-1/2}$, the decay rate of a mode with momentum $k$ into the non-commutative bath is significantly faster in comparison to the decay of the same mode into an ordinary commutative bath. 

It is interesting to note that in the regime where $k\gg \theta^{-1/2}$ the behavior of the leading dominant pole (as a function of temperature) is almost the same for the dilaton, the axion, and the minimally-coupled massless scalar field, which leads to a ``universal'' thermalization time in the UV.  This behavior could be attributed to the presence of the open Wilson lines in the definition of the gauge invariant operators in non-commutative gauge theories  which, as a result,  dominate the UV behavior of the two (and higher) point functions of the gauge invariant operators \cite{Gross:2000ba}. 

\section{Discussion}
\begin{figure}[h]
\centering
\hskip -0.1in \includegraphics[width=70mm]{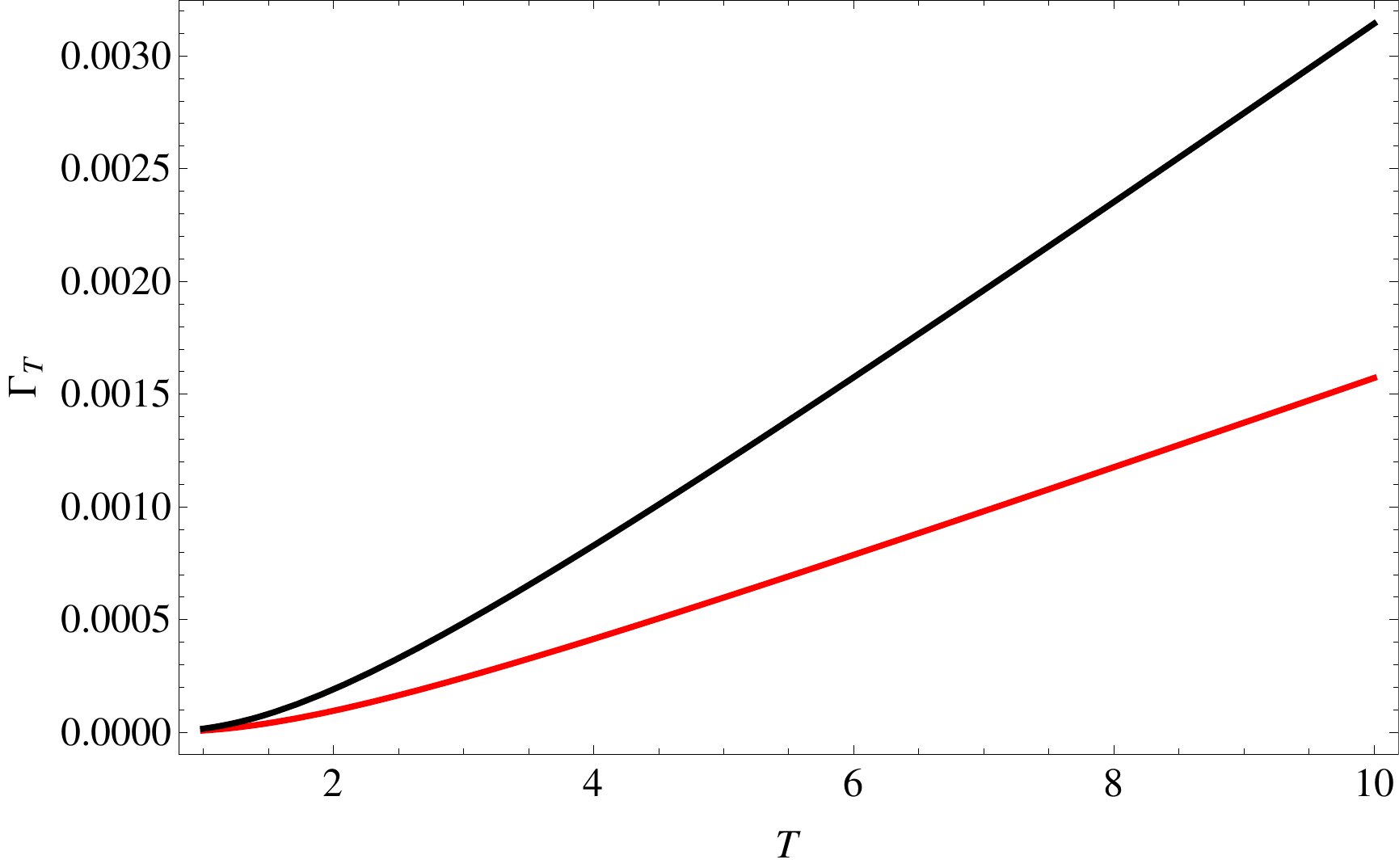}
\hskip 0.1in
\caption{\label{weak}\footnotesize{ Comparison of the thermalization
rates for the commutative (black) and noncommutative (red) cases at weak coupling.} }
\end{figure}

In this paper we showed that strongly coupled non-commutative gauge theories display at high temperature, a fast dissipation rate not seen in local field theory. At the core of this result is the UV-IR connection embodied in the large transverse size of high momenta mode.  
In order to assess how crucial the strong coupling ingredient is, it is useful to compare the behavior at high temperature in the weakly coupled regime. An example suggestive of this weakly coupled regime can be found in [36]. Although the model studied there is not a gauge theory, the leading perturbative contributions are similar to the ones found in gauge theories and therefore can be considered exemplary of the behavior at weak coupling. In that work, the authors computed the decay rate of a disturbance of the heat bath and found that the planar contributions to the thermalization rate are always larger than
or equal to the contributions from the non-planar
diagrams\footnote{Planar diagrams are the same as
in ordinary field theory. Non-planar diagrams have an extra factor
that depends on the Moyal tensor $e^{-\frac{i}{2}\,p_\mu
\theta^{\mu\nu} q_\nu}$. For a review of non-commutative quantum field
theories, see \cite{Minwalla:1999px, Douglas:2001ba}.}.  In
Figure (\ref{weak}) we show an example of such behavior for $k=10$ in
$1/\sqrt{\theta}$ units and small frequency $\omega$. It is clear from
the figure that the enhancement of the thermalization rates found here in 
strongly coupled non-commutative baths is not present in the corresponding weakly coupled theories.

It seems, then, that strong coupling is necessary to have faster
thermalization together with the non-locality inherent of non-commutative theories. We conclude that this
provides an important  ingredient for the study of fast scramblers. Indeed, there have been studies of the fast scrambling conjecture in
the framework of Matrix theory \cite{Sekino:2008he, Susskind:2011ap,
Asplund:2011qj}, which is known to be related to non-commutative theories \cite{Banks:1996vh, Connes:1997cr}. As a final remark, it would be interesting to investigate the decay rates in other non-local theories such as the theory dual to the near horizon limit of a stack of NS5-branes \cite{Aharony:1998ub}, the so-called little string theory, and compare the results with the findings in this paper.  

\section{Acknowledgments}
This material is based upon work supported by the National Science
Foundation under Grant Number PHY-0969020 and by the Texas Cosmology
Center. We are grateful to E. C\'aceres,  A. Kundu and O. Saremi for
helpful discussions.

\section*{Appendix}
In this appendix,  we show that the poles in the retarded Green's function of the boundary theory operator which is dual to a minimally-coupled massless scalar field in the bulk, are all in the lower half of the complex frequency plane. This then implies that the non-commutative field theory under consideration is stable against the small perturbations caused by turning on the aforementioned operator. Our analysis here follows the argument given in \cite{Horowitz:1999jd} where it was shown that the quasi-normal frequencies of a minimally-coupled massless scalar field in the Schwarzschild AdS background are all located in the lower half of the complex frequency plane. 

The equation of motion for a minimally-coupled massless scalar in the non-extremal background \eqref{NCbackground} is given in \eqref{SEoMMomentumSpace}.
Defining the tortoise coordinate $u_{*}$ by
\begin{align}
\frac{du_{*}}{du}= \frac{1}{u^2f(u)},
\end{align}
the equation \eqref{SEoMMomentumSpace} can be put in the form of a Schr\"{o}dinger equation
\begin{align}\label{SchrodingerEq}
\Psi''(u_{*})+\big[\omega^2-V(u_{*})\big]\Psi(u_{*})=0,
\end{align}
where $\Psi(u)=u^{3/2}\varphi(u)$ and 
\begin{align}
V(u_{*})=f(u)\left[\frac{k^2}{h(u)}+\frac{3}{2}u^3f'(u)+\frac{15}{4}u^2f(u)\right]. 
\end{align}
In the above expression, it is understood  that $u$ is a function of $u_{*}$. Note that the potential blows up in the asymptotic boundary ($u_{*}\to 0$) and vanishes exponentially at the horizon where $u_{*}\to -\infty$. 

Suppose now that $\Psi$ is a quasi-normal mode with the associated quasi-normal frequencies $\omega_n$. Since, by definition, the quasi-normal mode $\Psi$ is infalling near the horizon, we isolate its near horizon behavior and define
\begin{align}\label{Defpsi}
\Psi(u_{*})\sim e^{-i\omega u_{*}}\psi(u_{*}).
\end{align}
Note that $\psi(u_{*})$ vanishes in the asymptotic $u_{*}\to 0$ region. Substituting \eqref{Defpsi} into \eqref{SchrodingerEq}, and changing back to the $u$-coordinate, one easily obtains
\begin{align}\label{psiEq}
\hskip-0.075in\partial_u\left[u^2f(u)\partial_u\psi(u)\right]-2i\omega_n\partial_u\psi(u)+U(u)\psi(u)=0,
\end{align}
with
\begin{align}\label{DefU}
U(u)=\frac{k^2}{u^2h(u)}+\frac{3}{2}uf'(u)+\frac{15}{4}f(u).
\end{align}
Now, multiplying \eqref{psiEq} by $\bar\psi$ and integrating the result, we obtain
\begin{align}\label{IntegralpsiEq}
\int_1^\infty du\, \Big[&u^2f(u)|\psi'(u)|^2+2i\omega_n\bar\psi(u)\psi'(u)+U(u)|\psi(u)|^2\Big]=0,
\end{align}
where ewe have also performed an integration by parts. Subtracting \eqref{IntegralpsiEq} from its complex conjugate yields
\begin{align}
\int_1^\infty du\,\Big[\omega_n\bar\psi(u)\psi'(u)+\bar\omega_n\psi(u)\bar\psi'(u)\Big]=0,
\end{align}
which, after an integration by parts, results in
\begin{align}\label{OmegaEq}
({\rm Im}\,\omega_n)\int_1^\infty \bar\psi(u)\psi'(u) du=-\frac{i}{2}\bar\omega_n |\psi(u=1)|^2.
\end{align}
Note that, from the above equation, for ${\rm Re}\,\omega_n\neq 0$, one obtains  ${\rm Im}\, \omega_n\neq 0$. Substituting \eqref{OmegaEq} into \eqref{IntegralpsiEq}, one obtains
\begin{align}\label{ImOmegaEq}
&\int_1^\infty du\, \Big[u^2f(u)|\psi'(u)|^2+U(u)|\psi(u)|^2\Big]+\frac{1}{{\rm Im}\,\omega_n}|\omega_n|^2\,|\psi(u=1)|^2=0.
\end{align}
Since the potential $U(u)$, given in \eqref{DefU}, is positive definite for all (real) values of $k$ and $a$ in the range $u\in[1,\infty)$, one deduces from \eqref{ImOmegaEq} that ${\rm Im}\,\omega_n<0$.

\end{document}